\documentclass[aps,preprint]{revtex4}
\usepackage{graphicx}

\begin{document}

\draft
\title{Baryon spectra with instanton induced forces}
\author{C. Semay, F. Brau}
\address{Universit\'{e} de Mons-Hainaut, Place du Parc, 20,
B-7000 Mons, Belgium}
\author{B. Silvestre-Brac}
\address{Institut des Sciences Nucl\'{e}aires, IN2P3, CNRS, \\
Universit\'{e} Joseph Fourier, Av. des Martyrs, 53,
F-38026 Grenoble-Cedex, France}
\date{\today}

\begin{abstract}
Except the vibrational excitations of $K$ and $K^*$ mesons, the main
features of spectra of mesons composed of quarks $u$, $d$, and $s$ can
be quite well described by a semirelativistic potential model including
instanton induced forces. The spectra of baryons composed of the same
quarks is studied using the same model. The results and the
limitations of this approach are described. Some possible improvements
are suggested.
\end{abstract}


\maketitle

\section{Introduction}
\label{sec:intro}

The QCD based semirelativistic potential model is a successful approach
to describe both meson and baryon spectra. In most of these works, it is
assumed that the quark interaction is dominated by a linear confinement
potential, and that a residual interaction stems from the one-gluon
exchange mechanism. In particular, the spin-spin interaction implied by
this process is responsible of the non-degeneracy of $\pi$ and $\rho$
mesons. The results obtained with such models are generally in good
agreement with experimental data, but the mesons $\eta$ and $\eta'$
cannot
be described without adding an appropriate flavor mixing interaction.

\par Another QCD based candidate exists for the residual interaction:
The effective forces computed by 't Hooft from instanton effects
\cite{thoo76}. It is a pairing force which presents the peculiarities to
act only on quark-antiquark states with zero spin and zero angular
momentum in the nonrelativistic limit; it also generates constituent
masses for the light quarks. A
flavor mixing appears naturally with this interaction which has already
been used in various models to study light
mesons: Nonrelativistic potential model \cite{blas90}, instantaneous
Bethe-Salpeter
formalism \cite{munz94,rick00}, flux-tube model \cite{sema95},
semirelativistic
potential approach \cite{silv97,sema99}. In all cases, quite good
results are obtained.

\par A first attempt to test this
instanton induced forces for baryons is performed in Ref.~\cite{blas90}.
With the same set of parameters, a description of all light mesons and
baryons is obtained (only the constant potential is changed from mesons
to baryons). The ground states of spectra are quite well reproduced but
some meson and baryon excitations are obtained too high: Vibrational
excitations of $K$ and $K^*$ mesons, $\eta'$, $\phi$, vibrational
excitations of $N$, $\Lambda$, etc. Moreover, this model presents two
serious drawbacks. First, it is a nonrelativistic model. As the
velocity of a light quark inside a hadron is not small compared with the
speed of light, the interpretation of the parameters of the model is
questionable. Second, the constituent masses and the coupling constants
of the instanton induced forces have been considered has free parameters
fitted to reproduce at best meson spectra. Actually, these quantities
can be calculated from instanton theory.

\par More recently, an instantaneous Bethe-Salpeter three-body formalism
has been applied to the study of baryons with instanton induced forces
\cite{loer01a,loer01b,loer01c}. In these works, the three-body
generalization of the Hamiltonian developed in Ref.~\cite{rick00} is
used, but with parameters fitted for the baryons, that is to say
different from those found for the mesons. The constituent masses and
the coupling constants of the instanton induced forces are also
considered has free parameters. The results of these works are compared
with our results in the following.

\par In a previous work \cite{brau98}, we have developed  a
semirelativistic model for mesons including the instanton induced
forces, but with parameters calculated, as far as possible, with the
underlying theory. Very good results have been obtained with the
condition that the quarks are considered as effective degrees of freedom
with a finite size. In particular, all ground states of vector and
pseudoscalar mesons are well reproduced, generally better than in
Ref.~\cite{blas90}. The main flaw stems from the
usual problem of the vibrational excitations of $K$ and $K^*$ mesons. In
this paper, we use this model to describe baryons composed of
$u$, $d$, and $s$ quarks, in order to test the relevance of the
instanton induced forces in the framework of a semirelativistic
potential model.

\par Note that for some authors \cite{shif80} the pion should be treated
as a pseudo Goldstone boson and not as a quark-antiquark state.
Nevertheless, in
Refs.~\cite{blas90,munz94,rick00,sema95,silv97,sema99,brau98} devoted to
the study of mesons with instanton induced forces, the pion can always
be obtained with a correct mass. We think that this is not by chance
since, with the resulting pion wave function, it is possible to compute
the correct pion charge form factor \cite{brau02}, reasonable values for
the electromagnetic splittings \cite{brau98}, and (more convincing)
correct hadronic decay widths in which pions are produced \cite{bonn01}.
We believe that the instanton induced interaction can simulate processes
giving to the pion its very low mass, and that it is relevant to fix the
parameters of the Hamiltonian in order to obtain all pseudoscalar
mesons.

\par The main
characteristics of the model are recalled in
Sec.~\ref{sec:model}. The numerical technique used to compute the baryon
masses and the fitting procedures are briefly described in
Sec.~\ref{sec:res}, where various baryon spectra obtained are also
discussed. Some concluding remarks are given in Sec.~\ref{sec:rem}.

\section{Model}
\label{sec:model}

The model we use is the natural generalization to baryon of the
Hamiltonian built for mesons in Ref.~\cite{brau98}. It is worth noting
that this Hamiltonian is defined at the lowest order, that is to say
that none relativistic correction is included in the potential. Details
can be found in Ref.~\cite{brau98}.

\par The three-quark Hamiltonian is written
\begin{equation}
\label{hamil}
H= \sum_{i=1}^{3} \sqrt{{\vec p \,}_i^2 + m_i^2} + \sum_{i<j=1}^{3}
V_{ij},
\end{equation}
with $\vec p_i$ the momentum of quark $i$
($\sum_{i=1}^{3} \vec p_i = \vec 0$), $m_i$ its constituent mass,
and $V_{ij}$ the interaction between quarks $i$ and $j$.
The interaction contains the Cornell potential and the instanton
interaction. The Cornell potential, which depends only on distance
$r$ between two quarks, is given by
\begin{equation}
\label{cornell}
V_{\text{C}}(r) = \frac{1}{2} \left[ -\frac{\kappa}{r}+a\, r+C \right],
\end{equation}
where the $1/2$ factor takes into account the color contribution.
The confining part of this potential represents a good
approximation of the Y-shape string configuration.

The instanton induced interaction provides a suitable formalism to
reproduce well the spectrum of the pseudoscalar mesons and to explain
the masses  of the $\eta$- and $\eta'$-mesons.
In the nonrelativistic limit, this interaction between two quarks in
a baryon is written \cite{blas90,munz94}
\begin{equation}
\label{vinst}
V_{\text{I}}(r)=-4 \left(g P^{[nn]} + g' P^{[ns]} \right) P^{S=0}\delta(
\vec r\,),
\end{equation}
where $g$, $g'$ are two dimensioned constants, $P^{S=0}$ is the
projector
on spin 0, and $P^{[qq']}$ is the projector on antisymmetrical flavor
state $qq'$ ($n$ for $u$ or $d$ is a non-strange quark, and $s$ is the
strange quark). The operator $P^{[nn]}$ is simply
a projector on isosinglet states, but the operator $P^{[ns]}$ is not so
easy to implement. Indeed, the instanton interaction is obtained under
the hypothesis of a perfect SU(3) flavor symmetry \cite{blas90,munz94}.
So, the baryon wave function is assumed to have a definite spin-flavor
symmetry, as in the simple model of Ref.~\cite{shur89} used to calculate
baryon mass splitting. Within a more realistic model, the strange quark
is much heavier than a $n$-quark, and the wave function cannot have
a particular flavor symmetry other that an isospin symmetry for the
$n$-quarks. Consequently, the flavor matrix elements
$\langle ns |P^{[ns]}| ns \rangle$ have values in our model which are
half the values in Ref.~\cite{shur89}. To be compatible with this
reference, we have placed a supplementary factor 2 in front of the
operator $P^{[ns]}$ in the computation code. The procedure to handle the
same problem in Ref.~\cite{blas90} is not described.

\par The instanton induced forces also give a contribution $\Delta m_q$
to
the current quark mass $m^0_q$. As this interaction is not necessarily
the only source for the constituent mass, a phenomenological term
$\delta_q$ is also added to the current mass \cite{brau98}. Finally,
the constituent masses in our models are given by
\begin{eqnarray}
m_n &=& m_n^0+\Delta m_n+\delta_n, \\
m_s &=& m_s^0+\Delta m_s+\delta_s.
\end{eqnarray}
In the instanton theory, the quantities $g$, $g'$, $\Delta m_n$,
$\Delta m_s$ are given by integrals over the instanton size $\rho$ up to
a cutoff value $\rho_c$ (see for instance Ref.~\cite[formulas (5) to
(9)]{munz94}). These integrals can be
rewritten in a more interesting form for numerical calculations by
defining a dimensionless instanton size $x=\rho\Lambda$ where $\Lambda$
is the QCD scale parameter \cite{brau98}
\begin{eqnarray}
\label{ins12}
g&=&\frac{\delta\, \pi^2}{2}\frac{1}{\Lambda^3}\left[m^0_s\
\alpha_{11}(x_{\text{c}})-\frac{c_s}{\Lambda^2}
\ \alpha_{13}(x_{\text{c}})\right], \\
\label{ins13}
g'&=&\frac{\delta\, \pi^2}{2}\frac{1}{\Lambda^3}\left[m^0_n\
\alpha_{11}(x_{\text{c}})-\frac{c_n}{\Lambda^2}
\ \alpha_{13}(x_{\text{c}})\right], \\
\label{ins14}
\Delta m_n&=&\delta \frac{1}{\Lambda}\left[m^0_n m^0_s\
\alpha_{9}(x_{\text{c}})-\frac{(c_n
m^0_s+c_s m^0_n)}
{\Lambda^2}\ \alpha_{11}(x_{\text{c}})+\frac{c_n c_s}{\Lambda^4}\
\alpha_{13}(x_{\text{c}})\right], \\
\label{ins15}
\Delta m_s&=&\delta \frac{1}{\Lambda}\left[\left(m^0_n\right)^2\
\alpha_{9}(x_{\text{c}})-
2\frac{c_n m^0_n}{\Lambda^2}\
\alpha_{11}(x_{\text{c}})+\frac{\left(c_n\right)^2}{\Lambda^4}\
 \alpha_{13}(x_{\text{c}})\right],
\end{eqnarray}
with
\begin{equation}
\label{ins11}
\alpha_n(x_{\text{c}})=\int_0^{x_{\text{c}}} dx\, \left[9\,
\ln\left(\frac{1}{x}\right)+
\frac{32}{9}
\ln\left(\ln\left(\frac{1}{x}\right)\right)\right]^6\, x^n\,
\left(\ln\left(
\frac{1}{x}\right)\right)^{-32/9}.
\end{equation}
In these equations, $\delta=3.63\, 10^{-3} \times 4\pi^2/3$ and
$c_i=(2/3)
\pi^2 \langle \bar{q}_i q_i \rangle$ where
$\langle \bar{q}_i q_i \rangle$
is the quark condensate for the flavor $i$. Except
the quantity $x_c=\rho_c\Lambda$, all parameters involved in
Eqs.~(\ref{ins12})--(\ref{ins15}) have expected values from theoretical
and/or experimental considerations. The integration in Eq.~(\ref{ins11})
must be carried out until the ratio of the ln-term on the lnln-term into
the integral stays small \cite{blas90,brau98}. This ratio increases
with $x$ from zero at $x=x_1=1/e$ to very large values.
At $x=x_2 \approx 0.683105$, the value of this
ratio is 1. This last value corresponds to the minimum of the instanton
density (see Ref.~\cite[Fig.~1]{brau98}). Thus we define the parameter
$\epsilon$ by
\begin{equation}
\label{ins16}
\rho_c=x_c/\Lambda \quad \text{with} \quad
x_{\text{c}}=x_1+\epsilon(x_2-x_1) \quad \text{and} \quad \epsilon \in
[0,1].
\end{equation}
In this work $\epsilon$ is a pure phenomenological parameter whose value
must be comprised between 0 and 1. With this procedure, the value of the
cutoff instanton size in our model is comprised between 0.3 and 0.5 fm,
which is a reasonable range for this parameter \cite{blas90,loer01c}.

\par The quark masses used in our model are the constituent masses and
not
the current ones. It is then natural to suppose that a quark is not a
pure point-like particle, but an effective degree of freedom which is
dressed by the gluon and quark-antiquark pair clouds. The form that we
retain for the probability density of a quark is a Gaussian function
\begin{equation}
\label{rhoi}
\rho_i(\vec r\,) = \frac{1}{(\gamma_i\sqrt{\pi})^{3/2}} \exp(- r^2 /
\gamma_i^2).
\end{equation}
It is generally assumed that the quark size $\gamma_i$ depends on the
flavor. So, we consider two size parameters $\gamma_n$ and $\gamma_s$
for $n$ and $s$ quarks respectively. It is assumed that the dressed
expression $\widetilde O_{ij}(\vec r\,)$ of a bare operator $O_{ij}(\vec
r\,)$, which depends only on the relative distance $\vec r = \vec r_i -
\vec r_j$ between the quarks $q_i$ and $q_j$, is given by
\begin{equation}
\label{od}
\widetilde O_{ij}(\vec r\,) = \int d\vec r\,'\,O_{ij}(\vec r\,')
\rho_{ij}
(\vec r - \vec r\,'),
\end{equation}
where $\rho_{ij}$ is also a Gaussian function of type (\ref{rhoi})
with the size parameter $\gamma_{ij}$ given by
\begin{equation}
\label{gij}
\gamma_{ij}= \sqrt{\gamma_i^2 + \gamma_j^2}.
\end{equation}
This formula is chosen because the convolution of two Gaussian
functions,
with size parameters $\gamma_i$ and $\gamma_j$ respectively, is also a
Gaussian function with a size parameter given by Eq.~(\ref{gij}).

\par After convolution with the quark density, the Cornell dressed
potential has the following form
\begin{equation}
\label{vfund}
\widetilde V_{\text{C}}(r) = -\kappa \frac{\text{erf}(r/\gamma_{ij})}{r}
+ a\,r \left[\frac{\gamma_{ij}\,\exp(-r^2/\gamma_{ij}^2)}{\sqrt{\pi}
\,r} + \left( 1+ \frac{\gamma_{ij}^2}{2r^2} \right)
\text{erf}(r/\gamma_{ij}) \right] + C,
\end{equation}
while the Dirac-distribution in $V_{\text{I}}(r)$ is transformed into a
Gaussian function
\begin{equation}
\label{insgauss}
\widetilde\delta(\vec r\,) = \frac{1}{(\gamma_{ij}\sqrt{\pi})^3}
\exp(-r^2/\gamma_{ij}^2).
\end{equation}
Despite this convolution, we consider, for simplicity, that the
instanton induced forces act always only on $L=0$ states. Note that the
strange size quark can be vanishing provided the non-strange quark size
is non-zero. Indeed, $\gamma_s=0$ with $\gamma_n \ne 0$ yields
$\gamma_{nn} \ne 0$, $\gamma_{ns} \ne 0$; only $\gamma_{ss} = 0$. This
last value could pose a problem only in expression~(\ref{insgauss}). But
this situation never happens since the instanton interaction
$V_{\text{I}}$ is vanishing for a $ss$ pair (see Eq.~(\ref{vinst})). For
mesons, the situation is a little bit more complicated and it is
discussed in Ref.~\cite{brau98}.

\section{Numerical results}
\label{sec:res}

\subsection{Numerical technique}
\label{ssec:tech}

The eigenvalue equation is solved by developing the wave functions in
trial states built with harmonic oscillator states
$|nlm\rangle$. In
such a basis, the two-body matrix elements of the potential are
expressed in terms of the following quantities
\begin{equation}
\langle n'lm|V(r)|nlm\rangle=\sum_{p=l}^{l+n+n'} B(n',l,n,l,p)I_p,
\end{equation}
with
\begin{equation}
I_p=\frac{2}{\Gamma(p+3/2)}\int_0^{\infty} dx\ x^{2p+2} \exp(-x^2)
V(bx).
\end{equation}
The quantities $I_p$ are the Talmi's integrals, which depend on a
nonlinear parameter $b$, the oscillator length. The coefficients
$B(n',l,n,l,p)$ are geometric factors \cite{brod67} which can be
calculated once for all. To accelerate the convergence, we use two
oscillator lengths $b$, $b'$ in our basis. These two quantities are the
scale parameters of the two internal radial distances which can be
defined in a baryon. This method, which has originally been developed in
Ref.~\cite{nunb77} for nonrelativistic kinematics, works very well for
relativistic kinematics, as it is shown in Ref.~\cite{silv01}. The
details of
the technique used to calculate the matrix elements of the relativistic
kinetic energy operator can be also found in Ref.~\cite{silv01}.

\subsection{Fitting procedure}
\label{ssec:mini}

The purpose of this work is to extend to the baryons the results
obtained
for the meson spectra in Ref.~\cite{brau98}, and thus try to obtain a
satisfactory description of baryon spectra with a quite simple model.
Our approach is indeed very simple since we use only a spinless Salpeter
equation supplemented by a pure central potential and the
nonrelativistic limit of an instanton induced interaction, that is to
say that the potential is completely defined at the zero order of
quark speed. This is
sufficient to describe the bulk properties of mesons and baryons. Note
that the instanton induced interaction is essential to describe
pseudoscalar mesons and the baryon ground state properties, such as the
$N$--$\Delta$ splitting.

\par We need thirteen parameters to obtain a satisfactory spectrum for
the
mesons, and to be consistent, we keep the same set of parameters for the
baryons. This number could appear large, but some are strongly
constrained by theoretical or experimental considerations, while other
are unavoidable (see discussion in Ref.~\cite{brau98}). The instanton
interaction is defined by six parameters: The
current quark masses, the quark condensates for the flavors $n$ and $s$,
the QCD scale parameter $\Lambda$, and the maximum size
$\rho_{\text{c}}$ ($x_{\text{c}}/\Lambda$) of the instanton. Four other
parameters are
introduced: The effective sizes of the quarks $n$ and $s$, and two terms
$\delta_n$ and $\delta_s$, which contribute to the constituent quarks
masses.  It is worth
mentioning that, among the above parameters, $m^0_n$, $m^0_s$,
$\Lambda$, $c_n$,
$c_s$, $\delta_n$, $\delta_s$, and $\epsilon$ are intermediate
quantities used to compute the four parameters
$m_n$, $m_s$, $g$, and $g'$ which enter directly into the Hamiltonian.
Three unavoidable parameters are also used for the
central part of the potential: The slope of the confinement
$a$---for which reliable estimations exist---, the strength $\kappa$ for
the Coulomb-like part, and the constant
$C$ which renormalizes the masses. Consequently, we can say that only
six quantities are really free parameters (see Table~\ref{tab:param}).

\par To find the value of the parameters, we have minimized a $\chi^2$
function based on the masses of 11 well-known baryons
(see Table~\ref{tab:baryon})
\begin{equation}
\label{chi2}
\chi^2=\sum_i\left[\frac{M^{\text{th}}_i-M^{\text{exp}}_i}{\Delta
M^{\text{exp}}_i} \right]^2,
\end{equation}
where the quantity $\Delta M^{\text{exp}}_i$ is the error on the
experimental masses (it is fixed at the minimum value of 10 MeV, see
Ref.~\cite{brau98}).
To perform the minimization, we use the most recent version of the
MINUIT code from the CERN library \cite{jame75}.

\subsection{Baryon spectra}
\label{ssec:spectra}

We first compute baryon spectra with the parameters found in our
previous
paper for meson spectra~\cite{brau98}, but the results obtained are not
very good. For instance, the roper resonance is found 576 MeV above the
nucleon---which is not so bad---, but the $N$--$\Delta$ mass difference
computed is
212 MeV, which is much too small. This mass difference is generally
considered as a minimum requirement to be reproduced for a baryon model.

\par In a second step, we have searched for a set of parameters to
describe both baryon and meson spectra. All sets found are very similar
and present more or less the same qualities and the same flaws (best
results are obtained with the supplementary factor 2 in front of the
operator
$P^{[ns]}$). For
instance, in one of the best sets of parameters found, a good
$N$--$\Delta$
mass difference is obtained (280 MeV), but the Roper resonance
mass is then calculated around 150 MeV above its experimental value.
Moreover, if the meson spectra obtained do not differ significantly from
the ones found in our previous paper \cite{brau98}, two states are then
very badly described with respect to the others: The $\eta'$-meson is
found 36 MeV too high and the
$^3D_J$ states are computed 43 MeV too low. Despite a great number of
minimizations, we never succeeded to
find an ``acceptable'' set of parameters to describe satisfactorily both
meson and the baryon sectors.

\par
\begin{figure}
\includegraphics*[height=8cm]{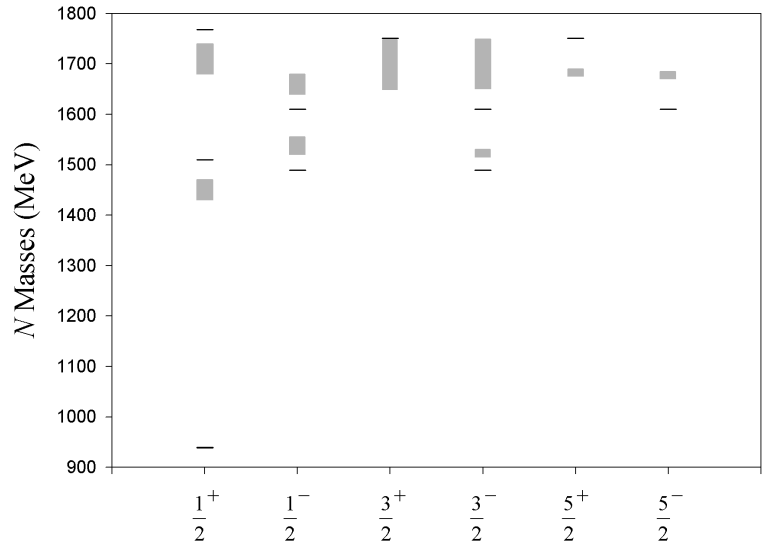}
\caption{Energy levels of the nucleon states (status
$\star$$\star$$\star$$\star$ and $\star$$\star$$\star$)
as a function of total
angular momentum and parity $J^P$. The shaded boxes represent the
experimental values with the uncertainties.}
\label{fig:n}
\end{figure}

\begin{figure}
\includegraphics*[height=8cm]{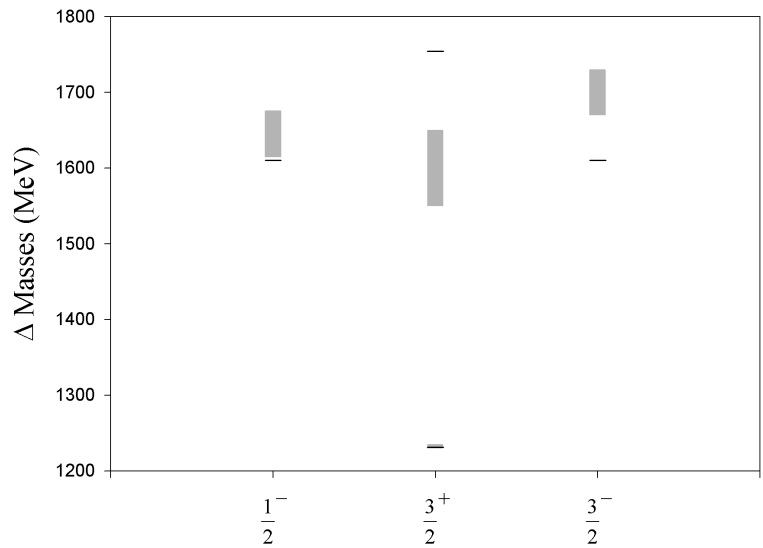}
\caption{Same as Fig.~\ref{fig:n} but for the $\Delta$ states.}
\label{fig:delta}
\end{figure}

\begin{figure}
\includegraphics*[height=8cm]{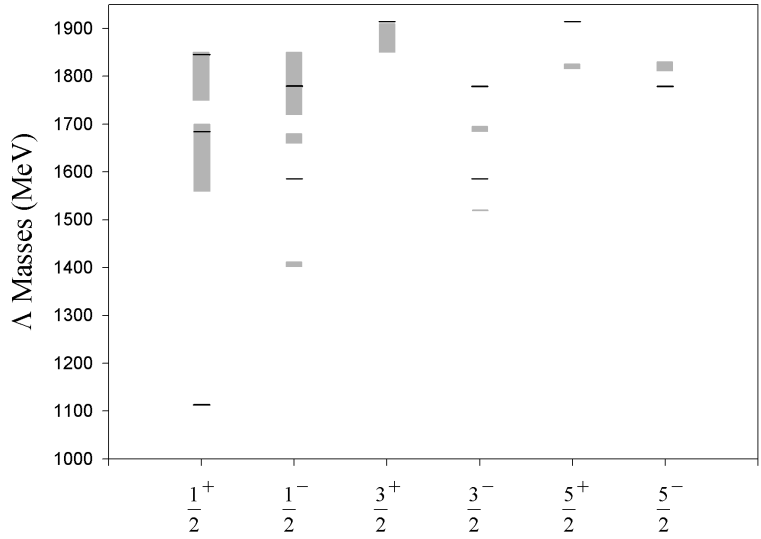}
\caption{Same as Fig.~\ref{fig:n} but for the $\Lambda$ states.}
\label{fig:lambda}
\end{figure}

\begin{figure}
\includegraphics*[height=8cm]{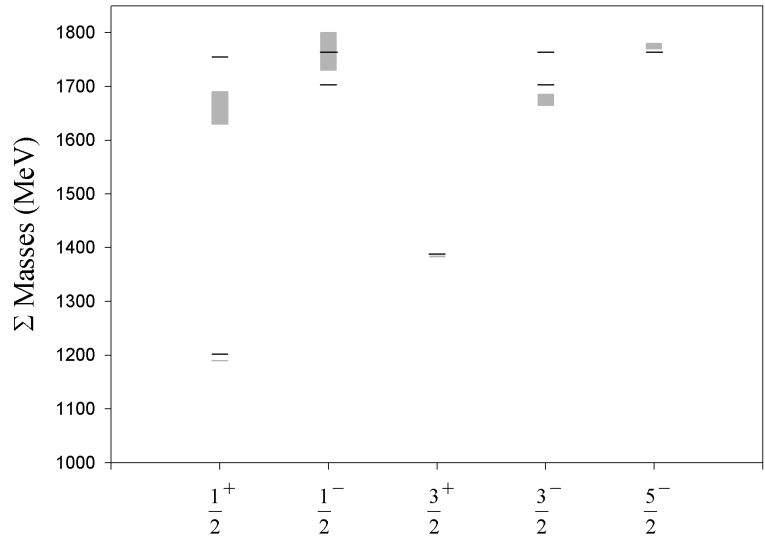}
\caption{Same as Fig.~\ref{fig:n} but for the $\Sigma$ states.}
\label{fig:sigma}
\end{figure}

\begin{figure}
\includegraphics*[height=8cm]{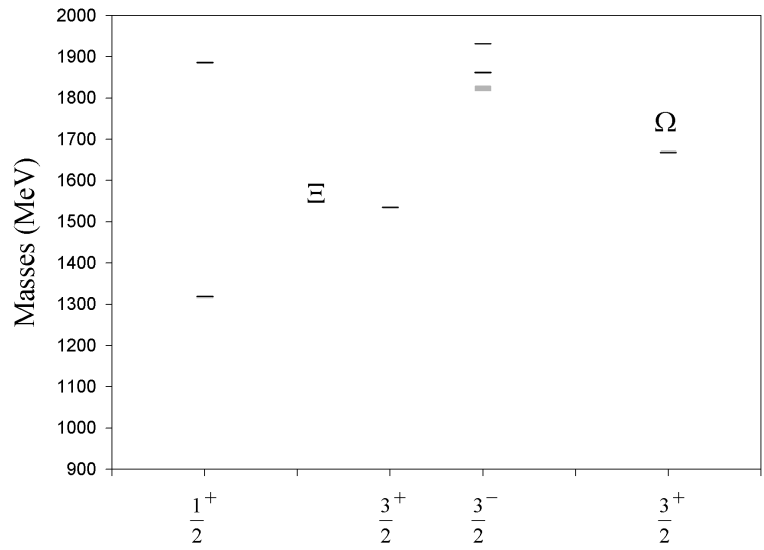}
\caption{Same as Fig.~\ref{fig:n} but for the $\Xi$ and $\Omega$
states.}
\label{fig:xi}
\end{figure}

 In order to test the relevance of our model for baryons, we have
then searched for parameters to describe baryons only. One of the best
set of baryon spectra that we have found is given in Fig.~\ref{fig:n} to
Fig.~\ref{fig:xi}. The spectra present some characteristics which can be
found in several other works, in particular Ref.~\cite{gloz98}; only few
states are not so well
reproduced in our work. For example, the mass of the Roper resonance is
around 60 MeV too high. The nucleon states with negative parity have
masses which are slightly too small. The Roper of the $\Delta$ has a too
high mass. The $\Lambda(1405)$ cannot be described, as this is often the
case. Even if our spectra are clearly less good than the spectra found
in Ref.~\cite{gloz98}, they presents many similar qualitative
characteristics. But in general, the agreement between calculated masses
and experimental data is less good in our model. It is worth noting that
all the baryon ground states can be well reproduced. With these
parameters fitted to the baryons, the mesons masses are very poorly
obtained: For instance, if
the computed pion mass is good (138 MeV), the mass of the $\rho$-meson
is found 260 MeV above its experimental value. The impossibility to
obtain good
meson and baryon spectra with the same parameters is also a
characteristic of the model of Ref.~\cite{gloz98}.

\par It is also interesting to compare our spectra with those obtained
in Refs.~\cite{loer01a,loer01b,loer01c}. The model developed in these
works and our model are similar in the sense that the instanton induced
interaction is the only spin-isospin-dependent part of the Hamiltonian,
but the model of Refs.~\cite{loer01a,loer01b,loer01c} differs from ours
by two main points: \emph{i)} the use of a spinless Salpeter equation in
our model instead of an instantaneous Bethe-Salpeter equation,
\emph{ii)} the presence of a Coulomb-like interaction in our model.
Below 2 GeV, spectra of both models are very similar; they share more or
less the same qualities and the same flaws: the ground states are well
reproduced, but the Roper resonance and the first $J^P=1/2^-$ state are
inverted. Again the $\Lambda(1405)$ cannot be described. We can just
note a slight improvement for other $\Lambda$-baryons. The baryon Regge
trajectories are nevertheless better described in the model of
Refs.~\cite{loer01a,loer01b,loer01c}; this is an indication of the
better relevance of an instantaneous Bethe-Salpeter equation over a
simpler spinless Salpeter equation. Note that our model is characterized
by smaller values of parameters $g$ and $g'$. This is due to the fact
that the instanton interaction can be weaker in our model since we
include in our Hamiltonian an attractive Coulomb-like interaction.

\par As one can see from Table~\ref{tab:param}, the values of the seven
first parameters are rather satisfactory ($\epsilon$ is expected to be
near zero, see Ref.~\cite{brau98}), while one can see that the
Coulomb-like parameter $\kappa$ is rather strong and that the size of
the $s$-quark is almost zero. A so small value for the strange quark
size could appear troublesome but this do not cause any numerical
difficulties, as mentioned above. Good meson and baryon spectra can be
computed with $\gamma_s$ around $0.5 \gamma_n$ and reasonable values
for the parameters $m_s^0$ and $\delta_s$. But the better baryon spectra
are obtained with small values of $\gamma_s$. As it is expected we found
$g' < g$
\cite{loer01b}. Note that when the parameters are fitted only
to baryons, the factor 2 in front of the operator $P^{[ns]}$ can be
simulated by a redefinition of the parameter $g'$. But in this case, we
have $g' > g$. It is also clear from
this Table that some of the best parameters for baryons are different
from the corresponding best parameters for mesons, in particular
quantities related to the quark masses. Moreover, the value of the
parameter $\kappa$ is higher for baryon, while the value of the
confinement slope $a$ is lower. With so much differences in these two
sets of parameters, it is not surprising that mesons and baryons cannot
be well reproduced together. Some physics is clearly missing. We discuss
this point in the next section.

\section{Concluding remarks}
\label{sec:rem}

Several works have been devoted to the study of the instanton
interaction in the framework of semirelativistic or relativistic models
for mesons
\cite{munz94,rick00,sema95,silv97,sema99,loer01a,loer01b,loer01c}. A few
of these
models have been applied to baryons. Our purpose was here to compute
baryon spectra with the semirelativistic instanton based model for
mesons we have developed in Ref.~\cite{brau98} and with the same
underlying fundamental ingredients.
When this model is directly applied to baryons, the spectra
obtained are not good. It is necessary to change all the parameters to
compute a more relevant spectra. The natural link between meson and
baryon is then broken, and the baryon spectra obtained are not very
different from
those yielded by models \emph{i)} with similar complexity but based on
one-gluon exchange process (see for instance Ref.~\cite{bhad81}),
\emph{ii)} relying on covariant equation with only an instanton induced
interaction supplementing the confinement
\cite{loer01a,loer01b,loer01c}.

\par Our semirelativistic model Hamiltonian contains a potential part
written in the lowest order, that is to say that none relativistic
correction is included. In particular, the spin-spin
term---responsible of the low pion mass in most potential models (see
for
instance Ref.~\cite{isgu86})---is not present here. The instanton
induced forces are assumed to take into account all spin effects. This
is probably a too crude approximation. Both interactions, instanton
induced
one and spin-spin term, have very similar contributions for non-mixed
flavor mesons. Nevertheless, to include the two interactions in a
Hamiltonian means a complete new fitting of parameters, in particular
the parameter $\kappa$ which measures the strength of the Coulomb-like
potential and of the spin-spin interaction. This could modify
appreciably
the spectra of baryons. It is not sure that the inclusion of
relativistic corrections in the model could cure all its defects. In
particular, the relative position of positive and negative parity
excitations of the nucleon is a problem for all models based on the
one-gluon exchange dominance. This puzzle is solved with the meson
exchange potential proposed more recently~\cite{gloz98}. Within this
model, the quarks interact by exchanging pseudoscalar mesons, completely
ruling out the one-gluon exchange process. Despite some serious critics
\cite{isgu00}, one is forced to ascertain that spectra of light baryons
are
remarkably improved. It is thus possible that the
meson exchange process could be one of the key to explain baryon spectra
and could supplement instanton induced interaction. Such a study is in
progress. Let us note that the Hamiltonian described in
Ref.~\cite{gloz98} cannot reproduce meson and baryon spectra with the
same set of parameters. It thus suffers the same drawback than our
model.

\par This work clearly shows that the instanton induced forces cannot
explain alone both meson and baryon spectra. Contributions coming from
one-gluon and meson exchange processes are probably necessary.
Nevertheless, as the 't Hooft interaction solves naturally the
$\pi$-$K$-$\eta$-$\eta'$ problem without any additional assumptions, it
must certainly be an unavoidable ingredient of potential models.

\acknowledgments

C. Semay would thank the F.N.R.S. for financial support. F. Brau
would thank the I.I.S.N. for financial support.


\clearpage

\begin{table}
\protect\caption{Quantum numbers and masses (the minimal uncertainty is
fixed at 10 MeV, see Ref.~\protect\cite{brau98}) of the baryons used in
the
minimization procedure to find the parameters listed in
Table~\protect\ref{tab:param}.}
\label{tab:baryon}
\begin{ruledtabular}
\begin{tabular}{lccr}
Baryon         & $I$           & $J^{P}$           & Masses (GeV) \\
\hline
$N$            & $\frac{1}{2}$ & $\frac{1}{2}^{+}$ & 0.939$\pm$0.010 \\
$N(1440)$      & $\frac{1}{2}$ & $\frac{1}{2}^{+}$ & 1.450$\pm$0.020 \\
$\Delta$       & $\frac{3}{2}$ & $\frac{3}{2}^{+}$ & 1.232$\pm$0.010 \\
$N(1535)$      & $\frac{1}{2}$ & $\frac{1}{2}^{-}$ & 1.537$\pm$0.018 \\
$\Lambda$      & 0             & $\frac{1}{2}^{+}$ & 1.116$\pm$0.010 \\
$\Sigma$       & 1             & $\frac{1}{2}^{+}$ & 1.193$\pm$0.010 \\
$\Sigma^*$     & 1             & $\frac{3}{2}^{+}$ & 1.385$\pm$0.010 \\
$\Xi$          & $\frac{1}{2}$ & $\frac{1}{2}^{+}$ & 1.315$\pm$0.010 \\
$\Xi^*$        & $\frac{1}{2}$ & $\frac{3}{2}^{+}$ & 1.530$\pm$0.010 \\
$\Omega$       & 0             & $\frac{3}{2}^{+}$ & 1.672$\pm$0.010 \\
$\Delta(1600)$ & $\frac{3}{2}$ & $\frac{3}{2}^{+}$ & 1.625$\pm$0.075
\end{tabular}
\end{ruledtabular}
\end{table}

\begin{table}
\protect\caption{List of parameters of the Model. The column
``Baryon'' contains the values for the baryon spectra presented in
Fig.~\protect\ref{fig:n} to Fig.~\protect\ref{fig:xi}. The column
``Meson'' contains the values for the meson model I of
Ref.~\protect\cite{brau98}. When available, the expected value of a
parameter is
also given in the column ``Exp.''. The values of the quantities $m_n$,
$m_s$, $g$, and $g'$ computed with these parameters are also indicated.}
\label{tab:param}
\begin{ruledtabular}
\begin{tabular}{llrrc}
Parameters                       & Unit        & Baryon & Meson & Exp.
\\
\hline
$m^0_n$                          & GeV       & 0.001 & 0.015 &
0.001--0.009 \protect\cite{pdg} \\
$m^0_s$                          & GeV       & 0.103 & 0.215 &
0.075--0.170 \protect\cite{pdg}\\
$\Lambda$                        & GeV       & 0.238 & 0.245 &
0.208$^{+0.025}_{-0.023}$\protect\cite{pdg} \\
$\langle\bar{n}n\rangle$         & GeV$^3$   & $-$(0.247)$^3$ &
$-$(0.243)$^3$ & $(-0.225\pm 0.025)^3$\protect\cite{rein85} \\
$\langle\bar{s}s\rangle/\langle\bar{n}n\rangle$ &  & 0.631 &
0.706 & $0.8 \pm 0.1$\cite{rein85} \\
$\epsilon$                       &           & 0.061 & 0.031 & 0--1
\protect\cite{brau98} \\
$a$                              & GeV$^2$   & 0.168 & 0.212 &
$0.20 \pm 0.03$ \cite{mich96} \\
$\kappa$                         &            & 0.798 & 0.440 & \\
$C$                              & GeV       & $-$0.967 & $-$0.666 & \\
$\gamma_{n}$                     & GeV$^{-1}$ & 0.681 & 0.736 & \\
$\gamma_{s}$                     & GeV$^{-1}$ & 0.005 & 0.515 & \\
$\delta_{n}$                     & GeV       & 0.327 & 0.120 & \\
$\delta_{s}$                     & GeV       & 0.490 & 0.173 & \\
\hline
$m_n$                            & GeV       & 0.378 & 0.192 & \\
$m_s$                            & GeV       & 0.638 & 0.420 & \\
$g$                              & GeV$^{-2}$   & 2.498& 2.743 & \\
$g'$                             & GeV$^{-2}$   & 2.234 & 1.571 & \\
\end{tabular}
\end{ruledtabular}
\end{table}

\clearpage

\end{document}